# The conductance of the quantum wire touching the gated Aharonov-Bohm ring

I.A. Shelykh[a)], N.T. Bagraev[b)], N.G. Galkin[b)], L.E. Klyachkin[b)]

[a)]School of Physics & Astronomy, University of Southampton, SO17 1BJ Southampton, UK
[b)]Ioffe Physico-Technical Institute, 194021 St.Petersburg, Russia

**Abstract.** We analyse the conductance of the Aharonov - Bohm (AB) one- dimensional quantum ring touching a quantum wire. The period of the AB oscillations is shown to be dependent strongly on the chemical potential and the Rashba coupling parameter that is in a good agreement with the studies of such a device prepared on the Si(100) surface.

## INTRODUCTION

The mesoscopic physics became the intense research field in last two decades. The gated AB rings present a special interest as they can be used as basic components for the realization of the spin interference devices [1]. The conductance of such a structure depends both on the magnetic and electric fields applied perpendicular to the plane of the AB ring. The former provides the AB phaseshift between the waves propagating in the clockwise and anticlockwise direction thus resulting in the conductance oscillations with the period in the range hc/2e – hc/e as a function of the amplitude of the backscattering on the contacts between the ring and the leads [2]. The electric field applied perpendicular to the plane of the ring provides the change of the carrier's wavenumber by shifting the subband's bottom inside the AB ring and by lifting the symmetry of the quantum well as a result of the Rashba spin - orbit interaction (SOI) [3]. The latter depends linearly on the gate voltage and creates the dynamical phaseshift between the waves propagating within the AB ring, which results in the Aharonov - Casher (AC) conductance oscillations [3].

In the present paper we analyze the conductance of the quantum ring touching the quantum wire that are narrow enough to support only one spin-degenerated propagating channel (Fig. 1). The drain-source voltage causing the electric current in the system is taken to be weak enough, so the Landauer-Buttiker formula can be used for the calculation of the conductance at zero temperature [4]:

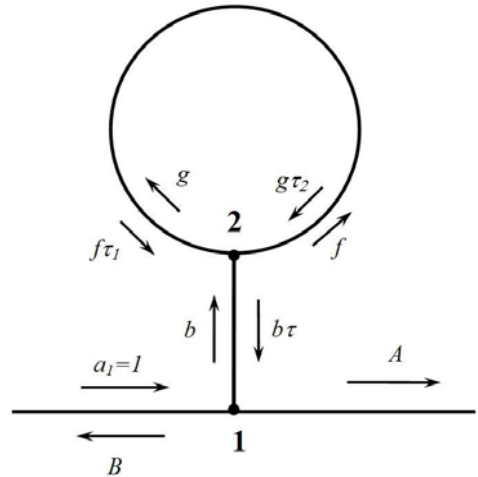

**FIGURE 1.** The amplitudes of the waves touching the quantum wire. The lead connecting the wire and the ring is introduced for the sake of the simplification of the calculations. Its length is then put equal to zero.

$$G = \frac{e^2}{h}\left[T_\uparrow(\Phi,\alpha,k_F) + T_\downarrow(\Phi,\alpha,k_F)\right] \quad (1)$$

where $T_{\uparrow,\downarrow}(\Phi,\alpha,k_F)$ is the transmission coefficient for the carriers of the two opposite spin directions dependent on the magnetic flux, $\Phi$, the Rashba coupling parameter, $\alpha$, and the Fermi wavenumber, $k_F$. We suppose also that the spin of the carrier is affected only by the Rashba SOI, because the Zeeman splitting can be neglected in weak magnetic fields.

## METHODS

In order to calculate the conductance of the device, we used the 3x3 scattering matrix, because the length of a short lead connecting the AB ring and the wire is supposed to be equal to zero. Within the adiabatic approximation the spin of the carrier follows the direction of the effective magnetic field created by the Rashba SOI, and the phase factors of the clockwise and anticlockwise propagating waves read

$$\tau_{1,2\uparrow} = \exp\left[i\left(2\pi k_{\pm}a \pm \frac{e\Phi}{\hbar c} \pm \theta_B\right)\right] \quad (2a)$$

$$\tau_{1,2\downarrow} = \exp\left[i\left(2\pi k_{\mp}a \pm \frac{e\Phi}{\hbar c} \mp \theta_B\right)\right] \quad (2b)$$

where the wavenumbers $k_{\pm}$ differ only if the Rashba SOI is taken into account and $\theta_B$ notes the Berry phase. [2].

## RESULTS

The conductance calculated as a function of $\alpha$, is seen to exhibit irregular oscillations determined by the variations of $k_{\pm}$ due to the AC effect and of $\theta_B$ with $\alpha$ if the external magnetic field is absent (Fig. 2a).

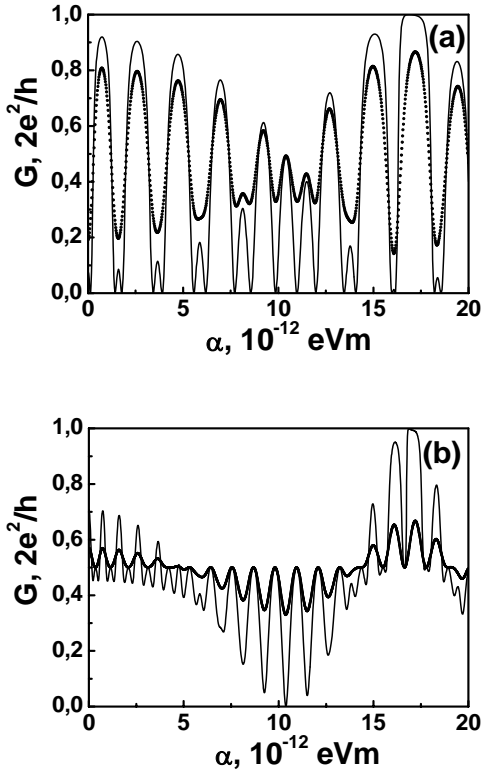

**FIGURE 2.** The dependence of the conductance on the Rashba SOI parameter for $\Phi/\Phi_0 = 0$ (a) and $\pi/2$ (b). $\mu = 10$ meV, T=0 and T=0.25 K. The radius of the ring is a = 0.5 μm.

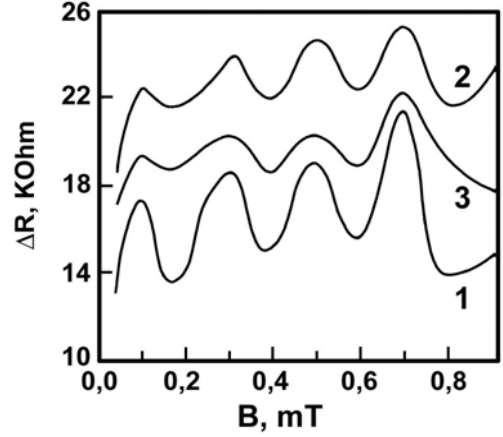

**FIGURE 3.** The AB conductance oscillations of the quantum wire touching the AB ring embedded in p-type Si-QW on the Si (100) surface that are observed by varying the Rashba SOI. T = 77 K;
α, $10^{-12}$ eVm: 1 – 2; 2 – 4.2; 3 – 9.5.

If the magnetic field and the Rashba SOI are present, the transmission amplitudes for the two spin components differ, because the transmitted and reflected currents become spin-polarized (Fig. 2b). The interplay between the AB and AC conductance oscillations shown in Figs. 2a and 2b appear to be revealed by studying the conductance of the quantum wire touching the AB ring that were electrostatically prepared inside the p-type Si-QW on the n-Si (100) surface (Fig. 3). The shape of the AB oscillations observed depends on the chemical potential and the Rashba SOI parameter, as it is predicted in Fig. 2b. In particular, these two quantities determine the relation between the amplitudes of the AB oscillations and the Aronov-Altshuler-Spivak oscillations of the half period [5].

## SUMMARY

The dependence of the conductance of the quantum wire touching the AB ring on the Rashba SOI parameter has been shown to define by both the geometrical Berry phase and the AC phase.